\documentclass[twocolumn, a4paper,10pt]{article}
\usepackage[top=2.5cm, bottom=2.5cm, left=2.0cm, right=2.0cm,
columnsep=0.8cm]{geometry}
\usepackage{enumitem}
\usepackage[hidelinks]{hyperref}
\usepackage{boxedminipage}
\usepackage{nopageno}
\usepackage{graphicx}
\usepackage{natbib}
\usepackage[font=it]{caption}
\usepackage[usenames,dvipsnames]{xcolor}
\usepackage{listings}
\usepackage{caption}
\usepackage{subcaption}

\usepackage{selinput}\SelectInputMappings{adieresis={ä},germandbls={ß}}
\usepackage{epsfig} 
\usepackage{times} 
\usepackage{amsmath,amssymb,amsfonts,bm,mathrsfs}
\usepackage[capitalise]{cleveref}
\usepackage{float}
\usepackage{tikz}
\usepackage[english]{babel}
\usepackage{algorithm}
\usepackage{algpseudocode}
\usepackage{verbatim}
\usepackage{makecell}
\usepackage{multirow}
\usepackage{textcomp,mathcomp}
\usepackage{soul}
\usepackage{listings}
\newcommand{\todo}[1]{{\color{black} #1}}
\newif\ifInitialSubmission
\InitialSubmissionfalse
\setlist{itemsep=-0.1cm,topsep=0.1cm,labelsep=0.3cm}
\setenumerate{leftmargin=*}
\makeatletter
\renewcommand\title[1]{\gdef\@title{\fontsize{12pt}{2pt}\bfseries{#1}}}
\makeatletter
\renewcommand\section{\@startsection{section}{1}{\z@}{3pt}{3pt}{\normalfont\large\bfseries}}
\makeatletter
\renewcommand\subsection{\@startsection{subsection}{1}{\z@}{\z@}{\z@}{\normalfont\normalsize\bfseries}}
\makeatletter
\renewcommand\subsection{\@startsection{subsection}{1}{\z@}{\z@}{0.1pt}{\normalfont\normalsize\bfseries}}

\title{%
Semi-automated Thermal Envelope Model Setup for Adaptive\\	
Model Predictive Control with Event-triggered System Identification } 	
\author{
	\ifInitialSubmission
		\phantom{Line 4} \\
		\phantom{Line 5} \\
		\phantom{Line 6} \\
		\phantom{Line 7} \\
		\phantom{Line 8} \\
		\phantom{Line 9}
	\else
		Lu Wan$^{1,2}$, Xiaobing Dai$^3$, Torsten Welfonder$^1$, Ekaterina Petrova$^2$, Pieter Pauwels$^2$ \\ 
        $^1$Robert Bosch GmbH, 71272 Renningen, Germany \\ 
		$^2$Group of Information Systems in the Built Environment, \\ 
        Eindhoven University of Technology, 
		5600 MB Eindhoven, Netherlands \\ 
		$^3$Chair of Information-oriented Control, Technical University of Munich, 80333 Munich, Germany \\ 
        \phantom{Line 9}
	\fi
}
\date{\vspace{-0.5cm}}	
\begin{document}

\maketitle

\section*{Abstract}	
\addtocounter{section}{1}
To reach carbon neutrality in the middle of this century, smart controls for building energy systems are urgently required. Model predictive control (MPC) demonstrates great potential in improving the performance of heating ventilation and air-conditioning (HVAC) systems, whereas its wide application in the building sector is impeded by the considerable manual efforts involved in setting up the control-oriented model. To facilitate the system identification (SI) of the building envelope as well as the configuration of the MPC algorithms with less human intervention, a semantic-assisted control framework is proposed in this paper. We first integrate different data sources required by the MPC algorithms such as the building topology, HVAC systems, sensor data stream and control settings in the form of a knowledge graph and then employ the data to set up the MPC algorithm automatically. Moreover, an event-triggered SI scheme is designed, to ensure the computational efficiency and accuracy of the MPC algorithm simultaneously. The proposed method is validated via simulations. The results demonstrate the practical relevance and effectiveness of the proposed semantics-assisted MPC framework with event-triggered learning of system dynamics.

\section*{Highlights}
\begin{itemize}
	\item Semantic web technologies
 	\item Ontology-based data integration
	\item Adaptive model predictive control
	\item Event-triggered system identification
\end{itemize}

\section*{Introduction}
The building sector takes up about $40\%$ of the primary energy consumption and the greenhouse gas emissions worldwide \citep{noauthor_building_2015}, more than half of which occurs during the operational stage. 
heating, ventilation and air-conditioning (HVAC) systems account for a large part of the building’s total energy use during the operation, inducing considerable CO$_2$ emission and monetary costs. Due to the increasing need for space cooling and heating globally, smart and economical control strategies are required for HVAC systems to decrease the carbon footprint. \looseness=-1

Model Predictive Control (MPC) is a control method that uses building models and disturbance forecasts to solve constrained optimization problems in a dynamic manner \citep{drgona_all_2020}. 
It has gained increasing attention in building control these years due to its effectiveness in energy cost reduction and energy efficiency improvement \citep{eydner2022real}.
The control-oriented prediction model for MPC includes the dynamic of the thermal envelope and HVAC systems.
In this study, we mainly deal with thermal envelope modeling, where Resistance and Capacitance (RC) models are widely employed \citep{atam2016control}.The lumped parameters of RC models, characterizing the heat transfer of buildings, are usually estimated through data-driven system identification (SI).
In recent studies, adaptive SI methods are often employed, which update the RC model parameters regularly, performing moving horizon estimation (MHE) or model identification at a daily frequency \citep{arroyo2020identification,blum2022field}. Because the performance of MPC deteriorates if the estimated RC parameters are inaccurate.
We adopt an MPC algorithm with event-triggered model identification, to reduce the computational effort caused by frequent SI.

The wide application of MPC algorithms is however a challenge in the building sector. Because the control-oriented model configuration requires the interpretation of different data sources such as building geometry, HVAC systems, and sensor measurement \citep{blum2022field}, demanding different domain experts' knowledge. Furthermore, the MPC algorithm requires forecasts on disturbances caused by e.g.weather and occupancy, and appropriate hyper-parameter setting e.g. prediction horizon. As buildings have miscellaneous geometry, building energy systems, and geological locations, such a model-based approach is labor-intensive and difficult to transfer among buildings. Most of the previous studies have focused solely on the automatic control model setup for optimal structure and parameters, but with little regard to the practical implementation efforts required by both SI and MPC algorithms.
For instance, \cite{de2016toolbox} has developed a python toolbox to identify the RC model for the building envelope, which is also deployed in \citep{arroyo2020identification} to automatically determine the optimal model structures for multi-zone buildings. However, the meta information required by the toolbox is manually configured.  
\cite{andriamamonjy2019automated} propose a tool-chain to generate RC model in Modelica, which automatically uses geometric data in the building information model (BIM) and monitoring data from the building management system (BMS). This study is insightful, but the connection between BIM model and BMS is established in a hard-coded manner and the implementation depends highly on the proprietary commercial software Revit. In \cite{blum2022field}, a comprehensive open-source toolchain is developed that can generate automatically the MPC algorithm (including both model parameter estimation and optimal control formulation).
Nevertheless, the data integration and interpretation process involved in the MPC algorithm generation is implemented specifically for the case study. Thus, the developed infrastructure is hard to transfer to another building. An integrated framework that collects and interprets the diverse data required by the MPC needs to be studied, in order to prompt the MPC applications. \looseness=-1

There is no mature common platform yet to keep the heterogeneous data well-connected in the architecture, engineering, and construction (AEC) industry. Semantic technologies are considered promising to solve the data silo dilemma \citep{pauwels2022buildings}. Many previous attempts use Industry Foundation Classes (IFC) \citep{borrmann2018industry}, which is an open BIM schema for data exchange in the AEC domain. However, IFC is not suitable for describing dynamic operational data such as sensor measurements, despite its strength in geometry modeling. Semantic web technologies (SWT) \footnote{\url{https://www.w3.org/standards/semanticweb/}}
enable the exchange and sharing of diverse data sources (semantic graphs) over the web, which is hard to achieve by using classical approaches.
Semantic graphs also referred to as knowledge graphs or metadata schema, contain structured information describing the meaning of the underlying data \citep{fierro_survey_nodate}. Built by ontologies that mean a specification of a conceptualization \citep{gruber1993translation}, semantic graphs are able to connect heterogeneous data. A few studies try to integrate the building data from the design and operation stages using SWT. \cite{mavrokapnidis2021linked} propose an SWT-based methodology to link static building design data modeled in IFC with dynamic sensor data modeled by Brick Schema \citep{balaji_brick_2018}. Building topology, product, and sensor data are connected using these two schemata for further exploitation. A similar approach is adopted in \citep{Chamari_Petrova_Pauwels_2022} to integrate the IFC data and BMS sensor data, by first converting different data sources into knowledge graphs and then linking them together. They build a vendor-neutral web application to visualize the 3D BIM model and the spatial-related sensor measurements in a common platform. \todo{The ontology-based data integration can enable more efficient data-driven applications via properly linked data sources. However, to the best of  the authors' knowledge, no existing study has investigated a semantics-aware framework for advanced building control, such as MPC.} \looseness=-1

In this paper, we propose a semantic-assisted control framework to support MPC applications in buildings. In the proposed framework, the data from diverse sources, including building geometry, building physics, and sensors, is first collected and managed in a machine-interpretable way and then employed to set up the algorithm automatically.
Furthermore, an MPC algorithm with event-triggered SI is designed and implemented, in order to minimize operating costs with desirable indoor temperatures.
The effectiveness of the proposed framework and algorithm is demonstrated through simulations.\looseness=-1

The paper is structured as follows. In the section \nameref{ch_system_modeling}, we introduce the building and the HVAC system. In the section \nameref{ch_architecture}, we explain the proposed control framework and elaborate on the designed MPC algorithm with event-triggered SI. The results section shows the effectiveness of the proposed approach via simulations. Finally, conclusions are drawn with remarks on future research.
\section*{System modeling} \label{ch_system_modeling}

In this paper, we study a typical system of a European office building, consisting of a one-zone building envelope equipped with a Variable Air Volume (VAV) flow system and a radiator heating system. 
The physical model of the system is adapted from \textit{``Buildings.Examples.ScalableBenchmarks.BuildingVAV.-One$\_$Floor$\_$OneZone"} provided in the open-source Modelica Buildings library \citep{wetter_modelica_2014}.
\todo{The existing control logic in the model is selected as the baseline, i.e., a rule-based controller (RBC) designed according to \cite{american_society_of_heating_sequences_2005}, and compared with the proposed MPC algorithm in the result section.}

\subsection*{Building model description}
The BESTEST Case 600 \citep{Standard_140}, a benchmark model for building energy simulation, is used in the study.
The envelope consists of a single zone with a window on the south facade and a constant infiltration mass flow rate. 
A BIM model for the building envelope is manually created.
The building represents a typical small-group office located in Stuttgart, Germany. Internal heat gains $q_{int}$ and occupancy table are set according to the local standard \citep{DIN_16798_1}.
Moreover, bounds for desired zone temperature $T_z$ are adjusted according to the occupancy. 
In particular, for the occupied time $t \!\in\! \mathbb{O}$ (8:00 to 18:00) during weekdays, the upper and lower bounds of the zone temperature are set as $T_{max}^{occ} \!=\! 27 ^o\!C$  and $T_{min}^{occ} \!=\! 21 ^o\!C$. 
For unoccupied time $t \!\notin\! \mathbb{O}$, the temperature bounds are relaxed to $T_{max\!}^{un\!} \!=\! 32 ^o\!C$ and $T_{min\!}^{un\!} \!=\! 17 ^o\!C$, respectively. 
The air in the thermal is assumed to be well-mixed.
\begin{figure}[htb]
	\centerline{\includegraphics[width=1.0\columnwidth]{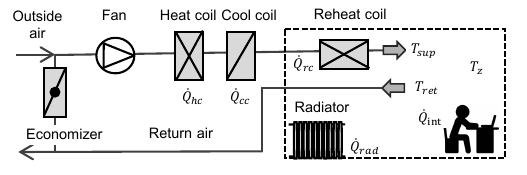}}
	\caption{
		Schematics of the HVAC system 
		}
	\label{figure_WholeSystem}
\end{figure}

The building is controlled by the system as in Figure \ref{figure_WholeSystem}, in which the VAV system consists of a heating coil, a cooling coil, and a reheat coil with maximal power $\dot{Q}^{hc}_{max}$, $\dot{Q}^{cc}_{max}$ and $\dot{Q}^{rc}_{max}$, respectively.
Moreover, an economizer is located between the main supply branch and the return branch. In the zone, the radiator heating system is deployed with a maximal power $\dot{Q}^{rad}_{max}$.
The HVAC components are modeled as ideal devices with constant overall efficiency, which considers the energy loss in the hydraulic distribution system and the generator system by using a discount coefficient.
The efficiencies are $\eta_{hc}$, $\eta_{rc}$,  $\eta_{rad}$ and ${COP}_{cc}$, respectively. 
The total thermal power delivered by the HVAC system $\dot{Q}^{hvac}$ to the zone is as in \cref{eqn_HVAC_power}:\looseness=-1
\begin{align} \label{eqn_HVAC_power}
	\dot{Q}^{hvac} = \bm{u}^T \bm{\Gamma} \dot{\bm{Q}}_{max}
\end{align}
where $\bm{u}^T=[u^{cc},u^{hc},u^{rc},u^{rad}]$ refer to the normalized power matrix of the component (control variable), with values between 0 and 1.
The efficiency matrix $\bm{\Gamma}$ is set as $\bm{\Gamma} = \mathrm{diag}({COP}_{cc}, \eta_{hc}, \eta_{rc}, \eta_{rad})$, while maximal power matrix set as $\dot{\bm{Q}}_{max}^T = [\dot{Q}^{cc}_{max}, \dot{Q}^{hc}_{max}, \dot{Q}^{rc}_{max}, \dot{Q}^{rad}_{max}]$.

\subsection*{Control problem setting}
The control objective is to achieve low operating costs while satisfying the requirements for the desired zone temperature. 
For this purpose, the MPC controller is designed with a first-order RC model for prediction. 
The continuous time 1R1C model is expressed as in \cref{eqn_RC}
\begin{align} \label{eqn_RC}
	C_z \dot{T}_z \!=\! R_{w}^{-1} (T_{amb} \!-\! T_z) \!+\! \dot{Q}^{hvac} \!+\! q_{int} A \!+\! \alpha H_{glo}
\end{align}
where $T_{amb}$ stands for the ambient temperature, $H_{glo}$ for the global horizontal irradiation, and $A$ for zone floor area. 
The parameter is set as $\bm{\theta} \!=\! [C_z,\! R_{w},\! \alpha]^T$, a collection of the heat capacity $C_z$, thermal resistance of the walls $R_{w}$ and solar irradiance coefficient $\alpha$.

The lumped parameter $\bm{\theta}$ is identified using data-driven methods with historical measurements as the training data.  
SI aims to find the parameters that minimize the difference between the true state and prediction, defined in \cref{eqn_SI}:\looseness=-1 
\begin{align} \label{eqn_SI}
	\hat{\bm{\theta}} \!=\! \arg\min_{\hat{\bm{\theta}}} J^s \!\!=\!\! \!\int\! \varepsilon^2(t,\hat{\bm{\theta}}) dt ~
	\mathrm{s.t.}~ \hat{\bm{\theta}} \in \bm{\Theta} = [\underline{\bm{\theta}}, \bar{\bm{\theta}}]
\end{align}
where $\varepsilon(t,\hat{\bm{\theta}})$ represents the deviation of RC model prediction compared with measured data.  
Note that for the optimization, the proper setup of initial guess $\hat{\bm{\theta}}_0$ and boundaries $\bm{\Theta}$ are necessary for plausible results.
As the real parameter $\bm{\theta}$ is time-varying, repeated SI is needed but induces more computational effort. To reduce the computational burden, we design the event-triggered SI strategy, updating the parameter via \cref{eqn_SI} only when the model error is large. The detail is discussed in section {\nameref{Control_Algorithms}}.

As SI described above and the MPC implementation afterward require both inputs from different sources, such as a proper setup of initial guess and continuous state measurements,  
this leads to the demand for an integrated data framework. 
We endeavor to integrate building design data (IFC file) and operational data (sensor measurements), in order to reduce the manual efforts required for the RC model set-up and the control algorithm. 
A semantic-assisted architecture is proposed, to realize the semi-automated setup of the proposed MPC algorithm with event-triggered SI.

\section*{Semantic-assisted architecture for MPC} \label{ch_architecture}
\begin{figure}[htb]
	\centerline{\includegraphics[width=0.83\columnwidth]{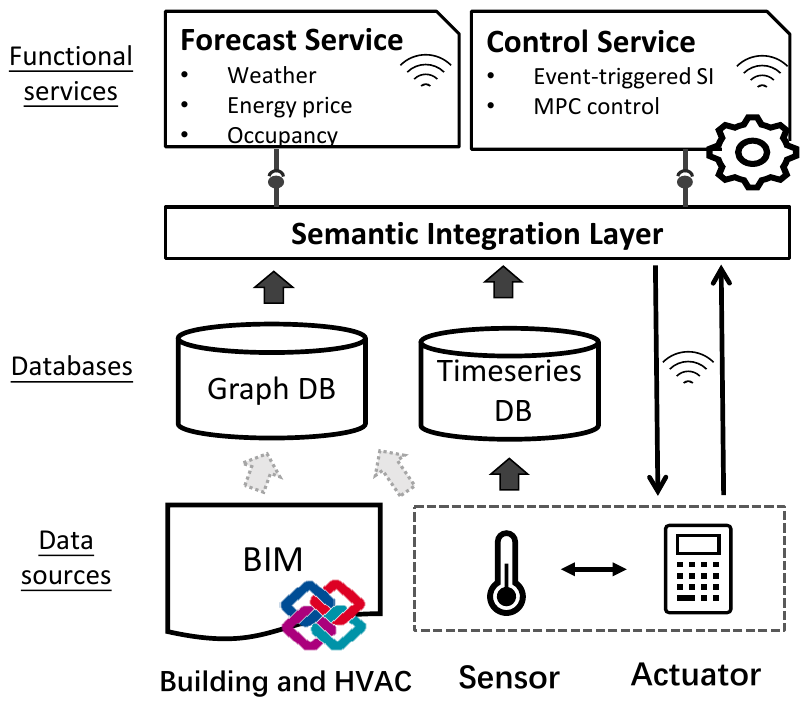}}
	\caption{
		The architecture of the semantic-assisted framework for MPC with even-triggered SI
	}
	\label{figure_Architecture}
\end{figure}
The proposed semantic-assisted framework adopts a layered service-oriented architecture as illustrated in Figure \ref{figure_Architecture}. 
Starting at the bottom, design data (BIM model) and operation data (data points of sensors and actuators) are first pre-processed and then delivered to the graph database and time-series databases accordingly. 
In the integration layer, semantic graphs for different chunks of entities e.g. buildings, HVAC systems, sensors and actuators are connected with each other and the overall information is integrated. The link between the virtual sensors in the graph DB and their measurements in the time-series DB is realized via the sensor ID. Via the semantic layer, the building design data and its operational data are seamlessly combined.
Eventually, the functional service layer communicates with the semantic integration layer, exchanging corresponding data to execute the services. Currently, there are two services implemented, namely the forecast and control services.

As for the software implementation, all the components except the MPC service are implemented in Python due to its versatile packages available. 
The manipulation of the semantic graphs is realized using RDFLib 
\footnote{\url{https://rdflib.readthedocs.io/en/stable/index.html}}. 
We use MATLAB to develop the MPC service because of its powerful numeric solvers available.

The real-time communications between the semantic layer and the databases, and between the field measurements and time-series database are via RESTful-APIs, which rely on HTTP protocol and are secure. 
The communication between the semantic layer and the services, as well as between the semantic layer and the field actuators, on the other hand, is realized via User Datagram Protocol (UDP), which is based on the TCP protocol and has the advantage of fast communication speed. 
In this way, the field devices (i.e. sensors and actuators) and the MPC service form a closed-loop system with a local area network.

Individual components are specifically explained in the following subsections.
\subsection*{Data sources and generation of graphs}
In this study, we take IFC file as the information source for data about envelope thermal properties, building geometry, and topology.
The semantic graph for building-related information is automatically generated by the IFCtoLBD convertor \citep{jyrki_oraskari_2023_7636217}. 
Since we use a simulation building with fictive HVAC systems, the graphs for the HVAC system and BMS data points are currently manually configured using Resource Description Framework (RDF) schema.
Note that the RDF graph generation for these sources can theoretically also be automated via tools described in \citep{Pauen:2022:FBI} and \citep{Chamari_Petrova_Pauwels_2022}, which will be handled in the future study.

\subsection*{Databases}
The generated semantic graphs for the building, HVAC system, and the metadata about sensors as well as actuators are stored in graph DB, because graph DB is fast in querying the relationships between entities. The schema used in the semantic graphs is detailed in the next section. On the contrary, the sensor measurements are stored in a time-series database, to enable more potential time-series analyses. We use GraphDB \footnote{\url{https://www.ontotext.com/products/graphdb/}} and InfluxDB \footnote{\url{https://www.influxdata.com/}} for the specific implementation.

\footnotetext[5]{
URIs for all used ontologies:\newline{}
BOT: \url{https://w3id.org/bot##};
Brick: \url{https://brickschema.org/schema/Brick##};
PEP: \url{https://w3id.org/pep/};
FSO: \url{https://w3id.org/fso##};
PROPS: \url{https://w3id.org/props##};
SEAS forecasting ontology: \url{https://w3id.org/seas/ForecastingOntology};
SOSA: \url{https://www.w3.org/ns/sosa/};
SSN: \url{https://www.w3.org/ns/ssn/};
TIME: \url{http://www.w3.org/2006/time/}.
}
\subsection*{Semantic integration layer} \label{ch_semantic_model}
In this subsection, we explain in detail how the semantic graph is modeled in terms of the adopted terminologies (T-box). This is fundamental for ontology-based data integration. Based on the controller design, we categorize the data and information that requires human inputs into the following five main aspects:

\begin{itemize} [leftmargin = 0.3cm, labelsep = 0.1cm, topsep = -0.1cm]
	\item \textbf{Building elements and topology.}
		The topological information and geometrical information about the thermal zones are required for RC model structure estimation and therefore modeled.
		In addition, the properties of building elements (walls, windows, doors etc.) such as area, thermal transmittance, solar transmittance and thermal capacitance are needed in calculating the initial guess and bounds of the parameter $\hat{\bm{\theta}}_0$.
		The information above is described using Building Topology Ontology (BOT)\footnotemark[5] \citep{rasmussen2021bot} and PROPS Ontology\footnotemark[5]  \citep{bonduel_towards_nodate}.
	\item \textbf{Components and topology for HVAC systems.}
		The properties of the HVAC system components, such as the nominal power $\dot{\bm{Q}}_{max}$, and the nominal efficiency $\bm{\Gamma}$ are modeled as \texttt{ssn:Properties}, using Semantic Sensor Network\footnotemark[5] (SSN) ontology \citep{compton2012ssn}.
		Furthermore, the interaction between the HVAC systems and zones is also modeled, so that the building envelope model is matched with the corresponding HVAC components efficiently. We use Flow System Ontology\footnotemark[5] (FSO) \citep{kukkonen2022ontology} to describe the heat and fluid transfer among zones and HVAC components.
	\item \textbf{Sensor data collection.} 
		The sensor data required by the MPC algorithm includes state measurement $T_z$. 
        In addition, the set-points i.e., $T_{max\!}^{(\cdot)}$ and $T_{min\!}^{(\cdot)}$, are required to setup the constraints in MPC as in \cref{eqn_MPC_T_constraint}. Such properties (e.g. temperature and power) observed by sensors are described by SOSA ontology\footnotemark[5] \citep{janowicz2019sosa}. The virtual data points attached to these properties are further modeled using Brick, with their identifier to the time-series measurements in the databases modeled via \texttt{brick:hasTimeseriesID}. 
	\item \textbf{Forecast information.}
		Forecasts required by the MPC algorithm include weather, occupancy, and the energy price in the market. 
        The link to the forecast models in the file system is described using SEAS forecasting ontology\footnotemark[5] and SEAS Procedure Execution ontology\footnotemark[5] \citep{lefranccois2017seas}. 
	\item \textbf{Controller setup.}
		The proposed adaptive control algorithm has 2 sub-modules as shown in Figure \ref{figure_Architecture}, namely MPC and the event-triggered SI. 
        For MPC, the prediction horizon $N_c$ is tuned according to the specific use case. For the event-triggered SI, the optimal setting for trigger horizon $N_t$ and identification horizon $N_s$, as well as the trigger threshold $\rho$ typically from domain experts’ experience. The above hyperparameters for controller setup are represented by SEAS optimization ontology and Time ontology\footnotemark[5] (Hobbs and Pan 2004), and can be modified by domain experts easily.
\end{itemize}
With the proposed information modeling paradigm, the information needed in building MPC algorithm is organized in a uniform manner, reducing the redundant efforts in data preparation. 
How the aforementioned information benefits the MPC algorithm setup is detailed in section \nameref{Control_Algorithms}. 
A concrete instance model (A-box) is shown in the section \nameref{A_box}.

The data from different sources are integrated and manipulated via the semantic integration layer, which serves as the coordinator between services and other components. The first task of the semantic integration layer is to calculate the initial guess $\hat{\bm{\theta}}_0$ and boundaries for RC parameters according to the material data, in which $\hat{R}_{w,0}$ is the sum of thermal resistance of all surfaces, and $\hat{C}_{z,0}$ for the total thermal capacitance of the surfaces and air within the zone, and $\hat{\alpha}_0$ for the solar irradiation through the window area. 
The lower and upper bounds of $\hat{\bm{\theta}}$ is set as $[0.1 \hat{\bm{\theta}}_0, 10 \hat{\bm{\theta}}_0]$.
The second task of the semantic layer is to retrieve the relevant data from the databases and delivery it to the corresponding service, which is detailed in the specific service. \looseness=-1

\subsection*{Functional service layer}
\subsubsection*{Forecast service}

\begin{figure}[hbt]
    \centerline{\includegraphics[width=1.0\columnwidth]{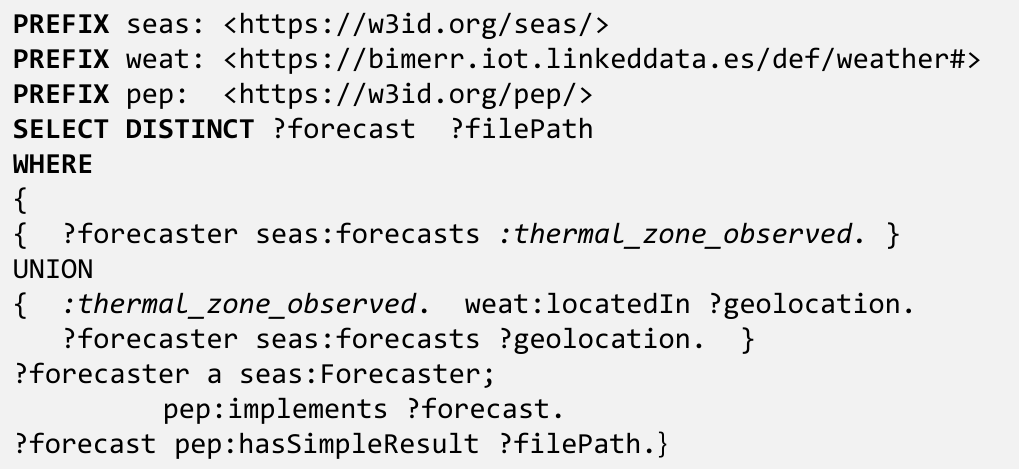}}
        \caption{SPARQL query in the forecast service to retrieve geological local related and zone-specific forecasts}
        \label{query_for_forecasts}
    \vspace{-0.5cm}
\end{figure}

The forecast service provides predictions on the weather environment, energy price and the occupancy of the building.
We use the predictions made by existing models and store them in the file system. Test reference year weather data set\footnote[6]{\url{https://energyplus.net/weather}} of Stuttgart is employed. We adopt the day-ahead German electricity price\footnote[7]{\url{https://www.smard.de/en/}} for price forecast. The occupancy profile as defined in the norm \citep{DIN_16798_1} is employed for predicting occupancy and internal gains. The SPARQL \footnote[8]{\url{https://www.w3.org/TR/sparql11-overview/}} query against prediction-related information is demonstrated in Figure \ref{query_for_forecasts}, running in the forecast service. \looseness=-1
\subsubsection*{MPC service with event-triggered SI} \label{Control_Algorithms}

The proposed algorithm in the control service consists of 2 sub-modules, namely MPC and event-triggered SI.

The MPC module is formulated as an optimization problem, aiming to minimize the operating costs of the HVAC system with desired zone temperature.
The optimization problem with receding horizon $N_c$ is written as \cref{eqn_MPC}:
\begin{subequations} \label{eqn_MPC}
	\begin{align}
		\min_{\{u_i,\bar{s}_i,\underline{s}_i\}}  \!\!\!\!\!\!\!&=\! J^c_k \!=\! \sum {_{i = k}^{k \!+\! N_c \!-\! 1}} \!( \hat{\lambda}_i \bm{\Gamma} \dot{\bm{Q}}_{max}^T \hat{\bm{u}}_i \!+\! \bar{\mu} \bar{s}_i \!+\! \underline{\mu} \underline{s}_i )
        \label{eqn_MPC_J} \\
		\mathrm{s.t.} ~ & T_{min}(t_i) - \underline{s}_i \le \hat{T}_{z, i} \le T_{max}(t_i) + \bar{s}_i \label{eqn_MPC_T_constraint}\\
		& \hat{u}^{cc\!}_i \!\in\!\! [0,\!1], \hat{u}^{hc\!}_i \!\in\!\! [0,\!1], \hat{u}^{rh\!}_i \!\in\!\! [0,\!1], \hat{u}^{rad\!}_i \!\in\!\! [0,\!1] \label{eqn_u_domain} \\
		& \underline{s}_i \ge 0, ~ \bar{s}_i \ge 0, ~ \forall i = k, \cdots, k + N_c \label{eqn_s_domain} \\
        & \hat{T}_{z, i + 1} = f(\hat{T}_{z,i},\! \hat{\bm{u}}_i,\! \hat{\bm{e}}_i,\! \hat{\bm{\theta}}_i), ~ \hat{T}_{z, k} = T_{z, k} \label{eqn_MPC_dynamics}
	\end{align}
\end{subequations}
where $\hat{\lambda}_i$ is the predicted electricity price, $\dot{\bm{Q}}_{max}$ is the maximal thermal power,and $\hat{u}_i$ is control variables.
The slack variables $\underline{s}_i$ and $\bar{s}_i$ stand for lower and upper bounds, and serve to guarantee the feasibility of the objective function by relaxing the hard constraints on $T_z$ as equation \cref{eqn_MPC_T_constraint}, with the penalty coefficient ${\mu}$ for violations.
\cref{eqn_MPC_T_constraint} and \cref{eqn_u_domain} are the constraints on the state ${T_{z,i}}$ and control variable $\bm{u_i}$, respectively. 
\cref{eqn_MPC_dynamics} is the discrete form of \cref{eqn_RC} with the estimated parameter $\hat{\bm{\theta}}_k$.

The estimated parameter is updated in an event-triggered way, i.e., $\hat{\bm{\theta}}_k \!=\! \gamma_k \hat{\bm{\theta}}_k^* \!+\! (1 \!-\! \gamma_k) \hat{\bm{\theta}}_{k\!-\!1}$, where the binary variable $\gamma_k$ indicates whether the parameter is updated. 
The newly estimated parameter $\hat{\bm{\theta}}_k^*$ is obtained from  historical data with length $N_s$ through the optimization as \cref{eqn_SI_discrete}:\looseness=-1
\begin{subequations} \label{eqn_SI_discrete}
	\begin{align}
		&\hat{\bm{\theta}}_k^* = \arg\min {_{\hat{\bm{\theta}}}} J^s_k = \sum {_{k - 1}^{k - N_s + 1}} \xi_i \varepsilon_i^2(\hat{\bm{\theta}}) \\
		&\mathrm{s.t.} ~  \hat{R}_{w} \in [\underline{R}, \bar{R}], ~ \hat{C}_z \in [\underline{C}, \bar{C}], ~ \hat{\alpha} \in [\underline{\alpha}, \bar{\alpha}]\\
		&\varepsilon_i(\hat{\bm{\theta}}) \!\!=\!\! T_{z,i \!+\! 1} \!\!-\!\! f\!(T_{z,i\!},\! \bm{u}_i,\! \bm{e}_i,\! \hat{\bm{\theta}}),\! \forall i \!\!=\!\! k \!\!-\!\! 1, \!\!\cdots\!\!, k \!\!-\!\! N_s \!\!+\!\! 1 \label{eqn_instant_model_error}
	\end{align}
\end{subequations}
where $\underline{R}$, $\underline{C}$, $\underline{\alpha}$ and $\bar{R}$, $\bar{C}$, $\bar{\alpha}$ are the constant lower and upper bounds of $R_{w}$, $C_z$ and $\alpha$.
The indicator $\gamma_k$ is the result from the event trigger, which is designed by considering $N_t$ previous data as \cref{eqn_trigger_condition}:
\begin{align} \label{eqn_trigger_condition}
    \mathrm{RMSE}_k \!=\! \sqrt{N_t^{-1} \sum {_{i \!=\! k \!-\! 1}^{k - N_t}} \varepsilon_i^2(\hat{\bm{\theta}}_k)} \!>\! \rho\Leftrightarrow \gamma_k \!=\! 1
\end{align}
where $\rho \!>\! 0$ is the trigger threshold.
The choice of $\rho$ is important but empirical, considering the trade-off between the trigger times and model-induced control accuracy.

According to the equations \eqref{eqn_MPC}, \eqref{eqn_SI_discrete} and \eqref{eqn_trigger_condition}, the setup of the MPC algorithm requires concrete data from (i) forecasts on electricity price, internal heat gains and ambient climate data, (ii) initial guess for the thermal envelope parameters and the boundaries, (iii) thermal zone and properties the HVAC components connected it, including nominal power and efficiency (iv) the specific control algorithm for the use case, e.g. horizons and threshold (v) the sensor measurements on states of the studied thermal zone, to which the semantic graph corresponds (described in section \nameref{ch_semantic_model}).

\begin{figure}[hbt]
        \centerline{\includegraphics[width=1.0\columnwidth]{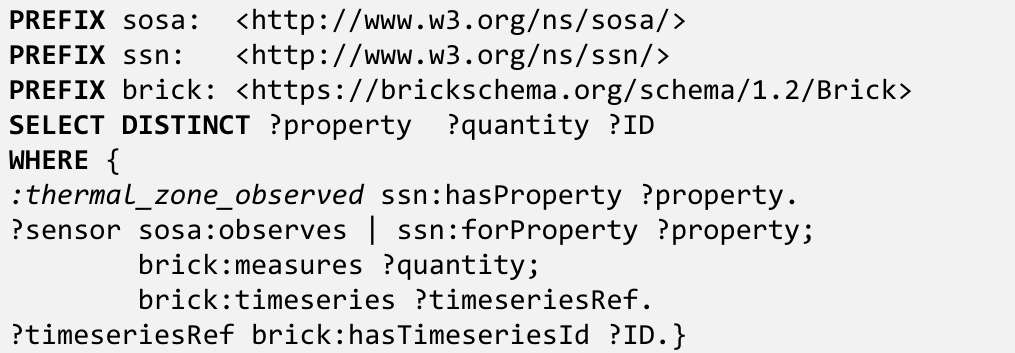}}
        \caption{SPARQL query for all sensor points of the zone
        }
        \label{figure_SPARQL_query_sensor}
\end{figure}

Regarding the sources, data (i) can be derived from the forecast service, data (ii) to (iv) come originally from BIM model and BMS and are stored in the graph DB, while data (v) is stored in the time-series DB. To fetch data (i)-(iv), SPARQL queries against the graph DB are sufficient. To retrieve data (v), the appropriate sensor ID needs to be firstly queried using SPARQL; afterward, the sensor ID as well as the time range need to be encoded into the FLUX query, to get the data out of InfluxDB. For illustration, the SPARQL query to retrieve all sensors related to the observed thermal zone is shown as in Figure \ref{figure_SPARQL_query_sensor}. 

With the assistance of the semantic integration layer and the underlying \nameref{ch_semantic_model}, the data required by the MPC service is easily redirected to the correct components and then merged to instantiate the algorithm depending on the specific use case. In this manner, the re-usability of the MPC among buildings is improved.

\section*{Results} \label{ch_Results}
\begin{figure*}[t]
	\centerline{\includegraphics[width=2\columnwidth]{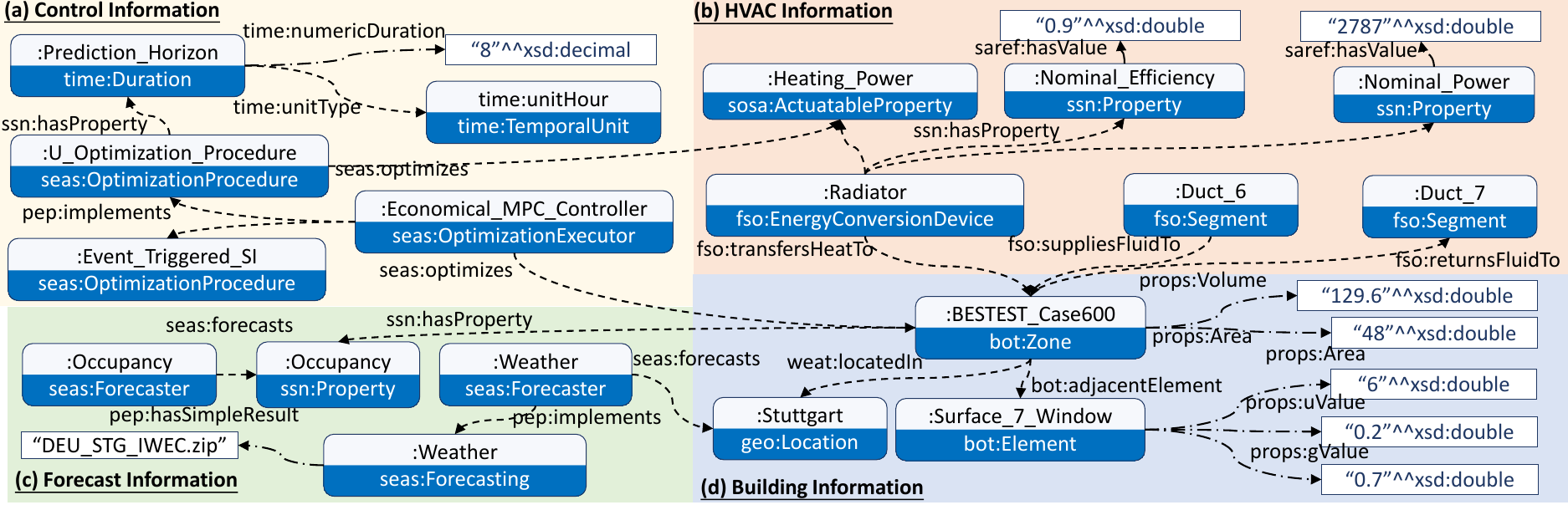}}
	\caption{
		Excerpt of the semantic graph for the studied case
	}
	\label{figure_SemanticModel}
	\vspace{-0.3cm}
\end{figure*}
\subsection*{Controller setup using the semantic graph} \label{A_box}
Here, we elaborate on how to set up the control algorithm with the semantic graph. An excerpt of the semantic graph deployed in the integration layer is shown in Figure \ref{figure_SemanticModel}. 

The lower right corner of Figure \ref{figure_SemanticModel} demonstrates the modeling of building-related information. The studied zone is modeled as a \texttt{bot:Zone} with a volume of $129.6\,m^3$ and an area of $48\,m^2$. It has 7 adjacent surfaces, including 4 walls, 1 floor, 1 ceiling, and 1 window. Taking the modeling of the window as an instance, the window (\texttt{bot:Element}) has an area of $6\,m^2$, a u-value of $0.2\,$$\frac{W}{m^2\cdot K}$$ $ and a g-Value of $0.7$, whereas its thermal capacitance is neglected. The properties of other surfaces are modeled in a similar way with one more property on the area-specific capacitance.
Using the data above, the initial guess and bounds of the RC parameters are calculated as $\bm{\hat{\theta}}_0\!=\![0.017\!\frac{K}{W},\!6.6\!\frac{MJ}{K},4.2\!m^2]$ and $\bm{\hat{\Theta}}_0\!=\![0.1\bm{\hat{\theta}}_0,\,10\bm{\hat{\theta}}_0]$, which are used by the event-triggered SI module in the MPC service.
Moreover, the information on sensor data points connected to the zone are queried using SPARQL described in Figure \ref{figure_SPARQL_query_sensor}, and the results are listed in Table \ref{table_sensor}. In total, four sensors are related to the zone (\texttt{:BESTEST$\_$case600}), measuring the  $T_{max}$, $T_{min}$, $T_z$ and occupancy head count. By using the sensor ID, the historic measurements made by the sensor are retrieved from time-series DB via Flux query. The historic measurements of $T_z$ are taken as an example and shown in Table \ref{table_sensor_temperature}. 
The historical measurements on occupancy and $T_z$ are sent to MPC service when SI procedure is triggered, while the real-time measurement of the state $T_z$ is sent to the MPC service every 5 minutes. Note the time-series historical weather data related to the building geological location (\texttt{:Stuttgart}) is sent to MPC service in a similar way, and is not detailed here. 

\begin{table}[htb]\small
	\caption{Sensor data points connected to the studied zone}
	\label{table_sensor}
	\centering
	\begin{tabular}{ c  c  c }
        \hline
		\bf{Property} & \bf{Quantity} & \bf{ID} \\
		\hline
		:Set$\_$Point$\_$Cool & brick:Temperature & LR101.TR22 \\
		:Set$\_$Point$\_$Heat & brick:Temperature & LR101.TR23 \\
		:Temperature & brick:Temperature & LR101.TR21 \\
		:Occupancy & brick:Occupancy$\_$Count & LR101.OC01 \\
        \hline
	\end{tabular}
\end{table}

\begin{table}[htb]\small
	\caption{Time-series data of zone temperature sensor (K)}
	\label{table_sensor_temperature}
	\centering
	\begin{tabular*}{0.9\columnwidth}{@{\extracolsep{\fill}}cc}
        \hline
		\bf{Time} & \bf{LR101.TR21} \\
		\hline
		2018-07-31 23:45:00 & 293.17 \\
		2018-07-31 23:50:00 & 293.14 \\
            2018-07-31 23:55:00 & 293.12 \\
        \hline
	\end{tabular*}
\end{table}

The right upper corner part illustrates the HVAC systems. 
\texttt{Duct$\_$6} and \texttt{Duct$\_$7} exchange fluids directly with the thermal zone, whereas the radiator exchanges heat directly with the zone. 
The radiator has nominal properties, such as $\dot{Q}^{rad}_{max}$ of 2787 W and $\eta_{rad}$ of $0.9$, and dynamic property $\dot{Q}^{rad}$, which is monitored and also optimized by the MPC optimization procedure. The rest of HVAC system's topology and their properties are modeled in a similar way. The maximal powers for the heating coil, reheat coil, and cooling coil are $1477\,W$, $261\,W$, and $-1814\,W$, with respective efficiencies 0.8, 0.8, and 2.7, where negative power means cooling. The maximal power matrix $\dot{\bm{Q}}_{max}$ and efficiency matrix $\bm{\Gamma}$ are utilized to initialize the MPC at the beginning, and historical measurements $\bm{\hat{Q}}$ are retrieved by the SI module when a system update is activated, in the same way as described in the last paragraph.

The forecaster information is modeled as the left lower corner of Figure \ref{figure_SemanticModel}. 
Using the query in Figure \ref{query_for_forecasts}, the forecast file paths are first retrieved via the forecast service (results in Table \ref{table_forecast}), and passed to the MPC service.
\begin{table}[htb]\small
	\caption{Results of the SPARQL query in forecast service}
	\label{table_forecast}
	\centering
	\begin{tabular*}{\columnwidth}{@{\extracolsep{\fill}}cc}
        \hline
		\bf{Forecast} & \bf{File} \\
		\hline
		:Day$\_$Ahead$\_$Electricity$\_$Forecast & Electricity$\_$STG.mat \\
		  :Occupancy$\_$Forecast & Occupancy$\_$Case$\_$600.mat \\
		:Weather$\_$Forecast & DEU$\_$Stuttgart$\_$IWEC.epw \\
        \hline
	\end{tabular*}
\end{table}

The information about the hyper-parameter settings in the MPC algorithm is modeled in the left upper corner of Figure \ref{figure_SemanticModel}.
The MPC service has two sub-modules: event-triggered SI and economic MPC. The former optimizes the RC model parameters to ensure the accuracy of the RC model, which is later used to predict the $\hat{T}_{z}$ in the MPC; the latter optimizes the relative power of all related components $\bm{u}$, to minimize the operating costs. For event-triggered SI, the trigger threshold $\rho$ is set as $0.1\,^o \! C$ with trigger horizon $N_t$ set as 1 day to ensure accurate daily prediction. The SI horizon $N_s$ (training data length) is set to 7 days, as recommended in \citep{blum2019practical, blum2022field}.
For the economic MPC module, we set the prediction horizon $N_c$ to 8 hours, because 1R1C model is not accurate for long-time prediction \citep{arroyo2020identification}. 
 
\subsection*{Performance of MPC with Event-triggered SI}
The simulation model in Modelica is exported as Functional Mockup Units (FMU), defined by Functional Mockup Interface (FMI) standard
, and simulated in Python with PyFMI library.
The simulation results of July are demonstrated as follows. \looseness=-1

\begin{figure}[htb]
	\centerline{\includegraphics[width=0.99\columnwidth]{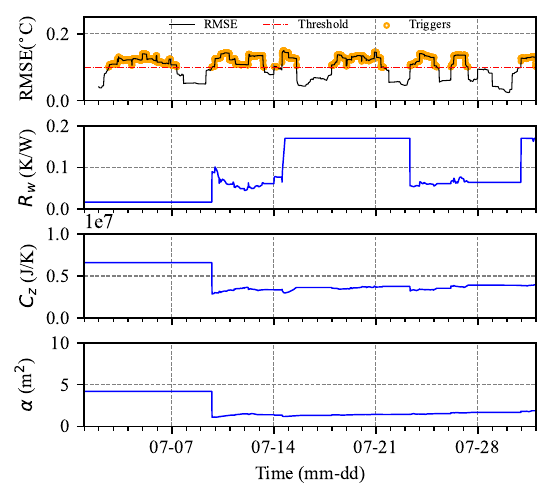}}
	\caption{
        Event-trigger and estimated parameters in July
	}
	\label{figure_ReultTrigger}
	\vspace{-0.3cm}
\end{figure}

Figure \ref{figure_ReultTrigger} shows the values of \todo{RMSE in \eqref{eqn_trigger_condition} with its threshold $\rho$} and the estimated parameters $\hat{\bm{\theta}}_k$ along the time in July. 
The evaluation of \todo{RMSE} starts after $1$ day, while the SI after $7$ days to collect enough data.
The SI is triggered repeatedly in certain periods because the eventual convergence of $\hat{\bm{\theta}}$ demands a few new data.
Moreover, more obvious changes are observed in $R_w$ than $C_z$ and $\alpha$, which results from their different sensitivities to the environmental boundaries and different effects on the short-time prediction ($8h$) accuracy.
Overall, a total of $3545$ times of SI is activated among $6912$ simulation steps by using the proposed event-trigger scheme, saving $49\%$ computations. \todo{Results compared to the MPC with MHE prove that the proposed event-triggered SI achieves similar control performance while demanding less computational effort.}
\begin{figure}[htb]
	\centerline{\includegraphics[width=0.99\columnwidth]{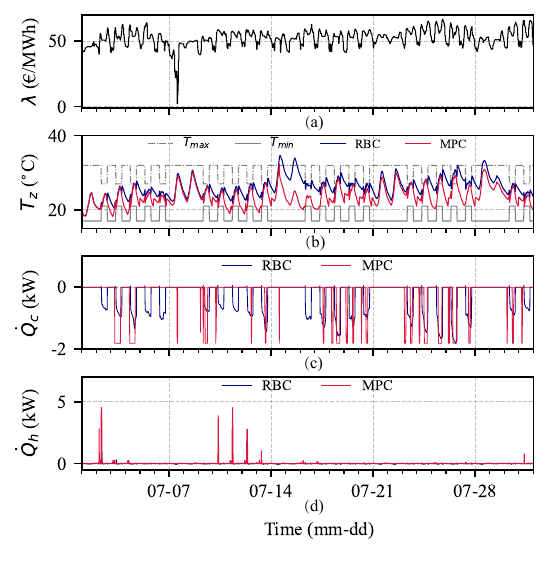}}
	\caption{
		Electricity price, zone temperature, cooling and heating power from HVAC system in July
	}
	\label{figure_ResultJuly}
	\vspace{-0.2cm}
\end{figure}

Figure \ref{figure_ResultJuly} shows the performance of the MPC algorithm in comparison to RBC model in July, which is defined in the system modeling section. 
More specifically, diagram (a) shows the varying electricity price, diagram (b) for the measured zone temperatures, diagram (c) for the total heating power $\dot{Q}^h \!=\! \dot{Q}^{hc} \!+\! \dot{Q}^{rh} \!+\! \dot{Q}^{rad}$, and diagram (5) for the cooling power $\dot{Q}^c \!=\! \dot{Q}^{cc}$.
According to the temperature profiles in Figure \ref{figure_ResultJuly}, the MPC algorithm controls the indoor climate better than the RBC, in terms of fewer violations of temperature constraints. 
Especially during the weekends (07-15,07-16 and 07-28), RBC model has forced the HVAC system to “Unoccupied-Off” mode, and no cooling power is supplied, while MPC algorithm predicts the temperature peak is about to come and provide moderate cooling power. 
Overall, MPC achieves an operating cost reduction of $12\%$ compared to RBC.

Combined with the successful integration of the building design data and sensor measurement explained, the results demonstrate the proposed semantic-assisted framework can be used practically to instantiate the algorithm.

\section*{Conclusion}
In this paper, a semantic-assisted control framework for MPC algorithm with event-triggered SI is proposed.
The framework facilitates MPC algorithm setup by integrating heterogeneous data sources via semantic modeling.
To ensure computational efficiency and the accuracy of MPC model at the same time, an event-triggered SI scheme is designed, where an educated initial guess and reasonable boundaries of RC model parameters for the thermal envelope are automatically instantiated using the semantic graph. 
The effectiveness of the proposed MPC algorithm is validated via simulations, where lower operating costs and better indoor temperature control are achieved, compared to the legacy RBC sequence. In the future study, the proposed framework will be verified on a real building.

\ifInitialSubmission

\else
\section*{Acknowledgment}
The authors would like to thank Dr. Philipp Kotman for the informative feedback dedicated to this paper. 
Xiaobing Dai is supported by the BMBF “Souverän. Digital. Vernetzt.” joint project 6G-life: 16KISK002. \looseness=-1
\fi

\bibliographystyle{bs2023}
\bibliography{references}

\end{document}